\documentclass[apj]{emulateapj}

\def\gtorder{\mathrel{\raise.3ex\hbox{$>$}\mkern-14mu
             \lower0.6ex\hbox{$\sim$}}}
\def\ltorder{\mathrel{\raise.3ex\hbox{$<$}\mkern-14mu
             \lower0.6ex\hbox{$\sim$}}}

\usepackage{colordvi}
\usepackage{amsmath}

\slugcomment{Draft of \today}

\shorttitle{AT\,2018lqh}

\shortauthors{Ofek et al.}

\begin{document}

\title{AT\,2018lqh and the nature of the emerging population of day-scale duration optical transients}
\author{E.~O.~Ofek\altaffilmark{1},
S.~M.~Adams\altaffilmark{2},
E.~Waxman\altaffilmark{1},
A.~Sharon\altaffilmark{1},
D.~Kushnir\altaffilmark{1},
A. Horesh\altaffilmark{1},
A.~Ho\altaffilmark{2},
M.~M.~Kasliwal\altaffilmark{2},
O.~Yaron\altaffilmark{1},
A.~Gal-Yam\altaffilmark{1},
S.~R.~Kulkarni\altaffilmark{2},
E.~Bellm\altaffilmark{3},
F.~Masci\altaffilmark{4},
D.~Shupe\altaffilmark{4},
R.~Dekany\altaffilmark{2},
M.~Graham\altaffilmark{2},
R.~Riddle\altaffilmark{2},
D.~Duev\altaffilmark{2},
I.~Andreoni\altaffilmark{2},
A.~Mahabal\altaffilmark{2},
A.~Drake\altaffilmark{2},
}

\altaffiltext{1}{Department of particle physics and astrophysics, Weizmann Institute of Science, 76100 Rehovot, Israel.}
\altaffiltext{2}{Cahill Center for Astronomy and Astrophysics, California Institute of Technology, Pasadena, CA 91125, USA}
\altaffiltext{3}{DIRAC Institute, Department of Astronomy, University of Washington, 3910 15th Ave. NE, Seattle, WA 98195, USA}
\altaffiltext{4}{Infrared Processing and Analysis Center, California Institute of Technology, Pasadena, CA 91125, USA}


\begin{abstract}


We report on the discovery of AT\,2018lqh (ZTF\,18abfzgpl) -- a rapidly-evolving extra-galactic
transient in a star-forming host at 242\,Mpc.
The transient $g$-band light curve's duration above half-maximum light is
about 2.1\,days, where 0.4/1.7\,days are spent on the rise/decay, respectively.
The estimated bolometric light curve of this object peaked at about $7\times10^{42}$\,erg\,s$^{-1}$ --
roughly seven times brighter than the Neutron Star (NS)-NS merger event AT\,2017gfo.
We show that this event can be explained by an explosion with a fast ($v\sim0.08$\,c) low-mass ($\approx0.07$\,M$_{\odot}$) ejecta, composed mostly of radioactive elements.
For example, 
ejecta dominated by $^{56}$Ni with a time scale
of $t_{0}\cong1.6$\,days for the ejecta to become optically thin for $\gamma$-rays
fits the data well.
Such a scenario requires burning at densities that are typically found in the envelopes of neutron stars or the cores of white dwarfs.
A combination of circumstellar material (CSM) interaction power at early times and 
shock cooling at late times is consistent
with the photometric observations, but 
the observed spectrum of the event may pose some challenges for this scenario.
We argue that the observations are not consistent with a shock breakout from a stellar envelope,
while a model involving a low-mass ejecta ramming into low-mass CSM cannot explain both the early-
and late-time observations.

\end{abstract}


\section{Introduction}
\label{sec:Introduction}

With sky surveys covering more volume per unit time,
rarer and rapidly evolving (i.e., fast rise, fast decline) transients are being discovered.
\cite{Drout+2014_Rapidly_Evolving}, \cite{Arcavi+2016_FastTransients} and, more recently,
\cite{Pursiainen+2018_DES_FastTransients} and \cite{Ho+2021_ZTF_RapidlyEvolvingTransients} reported several examples of transients
with a duration above half-maximum light of $\sim3$ to 12\,days.
Additional examples include, among others, PTF\,09uj \citep{Ofek+2010_PTF09uj_windBreakout},
AT\,2017gfo \citep{Abbott+2017_GW170817_MultiMessengerObservations}, AT\,2018cow \citep{Perley+2019_SN2018cow}
and AT\,2020xnd (\citealt{Perley+2020_AT2020xnd_FastLuminousTransient}).
These transients span a wide range of peak absolute magnitudes, from $-16$ to $-21$.

Transients featuring a fast evolution
usually probe physical explosions with low-mass ejecta or
high expansion velocities, or explosions embedded in low-mass optically-thick circumstellar material (CSM).
Among the astrophysical events that are expected
to produce such explosions are neutron stars' merger optical afterglows (e.g., \citealt{Abbott+2017_GW170817_MultiMessengerObservations}),
accretion-induced collapse (e.g., \citealt{Dessart+2006_AIC}; \citealt{Lyutikov+2018_ElectronCaptureCollapse}; \citealt{Sharon+Kushnir2019_AIC}),
shock breakout in an optically thick wind environment (e.g., \citealt{Ofek+2010_PTF09uj_windBreakout})
and, possibly, failed supernova (SN) explosions (e.g., \citealt{Adams+2017_FailedSN_search}; \citealt{Adams+2017_FailedSN_Candidate}; \citealt{Quataert+2019_FailedSN_BHaccretion}; \citealt{Fernandez+2018_FailedSN_massejection})
and  pulsational pair instability SN (e.g., \citealt{Barkat+1967_PISN}; \citealt{Waldman2008_PISN}; \citealt{Wooseley2017_PISN}).

Here we report on the discovery and followup of AT\,2018lqh -- a $7\times10^{42}$\,erg\,s$^{-1}$-peak luminosity,
fast-evolving transient found by the Zwicky Transient Facility (ZTF).
This object is among the shortest extragalactic transients
discovered by optical telescopes.
We present the observations and discuss the nature of this transient.
In \S\ref{sec:Obs}, we present the discovery and the observations,
while in \S\ref{sec:disc}, we discuss the physical nature of the event.
We conclude in \S\ref{sec:summary}.

\section{Observations}
\label{sec:Obs}

The Zwicky Transient Facility (\citealt{Bellm+2019_ZTF_Overview}, \citealt{Graham+2019_ZTF_objectives}, \citealt{Dekany+2020_ZTF_Camera}) is a sky survey utilizing
the 48-inch (P48) Schmidt telescope
on Mount Palomar, equipped with a 47\,deg$^{2}$ camera.
One its programs entailed the scanning of
about 1500\,deg$^{2}$, at least three times per night.
The scheduler is described in \cite{Bellm+2019_ZTF_Scheduler},
and the machine learning process responsible for identifying
the most-probable transients from the many image artifacts
is discussed in \cite{Duev+2019_RealBogousZTF}, while
the alerts screening and filtering is discussed in \cite{Nordin+2019_AMPEL} and uses the tools
in \cite{Soumagnac+Ofek2018_catsHTM}
and the ZTF marshal (\citealt{Kasliwal+2019_GROWTH_Marshal}).
The ZTF project also utilizes the Palomar 60-inch (P60) telescope
(\citealt{Cenko+2006_P60_Overview}), equipped
with the Spectroscopic Energy Distribution Machine and
Rainbow camera (\citealt{Ben-Ami+2012_SEDM}; \citealt{Blagorodnova+2018_SEDM}).

AT\,2018lqh was first automatically detected by ZTF on 2018 Jul 12 (JD 2458311.6850)
at J2000.0 coordinates of: 	$\alpha=16^{h}06^{m}04.^{s}47$, $\delta=+36^{\circ}52'16.''5$
($\alpha=241.518638^{\circ}$, $\delta=+36.871243^{\circ}$).
The event took place in a star-forming galaxy,
and the Sloan Digital Sky Survey (SDSS; \citealt{York+2000_SDSS_Summary}) spectrum of this
galaxy shows narrow Balmer, \ion{N}{2}, and \ion{O}{2} emission lines at $z=0.05446$ (242\,Mpc).

Throughout the paper, we assume the light curve and spectrum
were affected by a Galactic extinction with a
reddening of $E_{B-V}=0.019$\,mag (\citealt{Schlegel+1998_DustMapsReddening})
and assume an extinction law with
$R_{V}=3.08$ (\citealt{Cardelli+1989_Extinction}; see \S\ref{sec:spec}).

Unless specified otherwise, the analysis was performed using tools available in \cite{Ofek2014_MAAT}, with
all magnitudes given in the AB system.

\subsection{Photometry}
\label{sec:phot}

\begin{figure}
\centerline{\includegraphics[width=8cm]{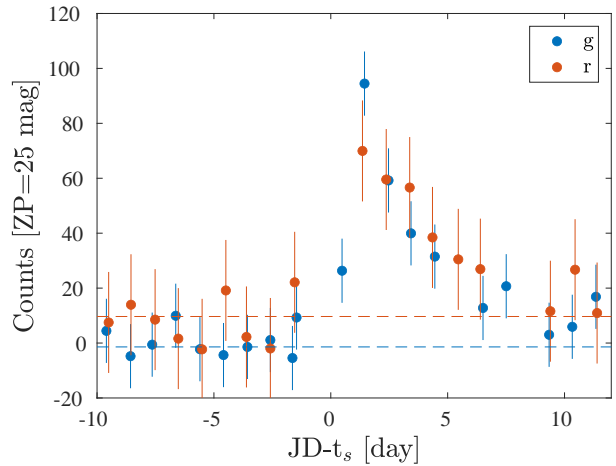}}
\caption{The $g$- (blue) and $r$-band (orange) ZTF nightly binned observations around the transient discovery. Time is measured relative to $t_{\rm s}=2458310.348$.
The dashed lines represent the mean level of the flux residuals
between 15 and 5 days prior to $t_{\rm s}$.
The error bars represent statistical errors, while
the mean level of the flux residuals is due to some
systematics.
{\bf This systematic does~not have any impact on the main results presented in this work.}
\label{fig:ZTF18gpl_gr}}
\end{figure}

\begin{figure}
\centerline{\includegraphics[width=8cm]{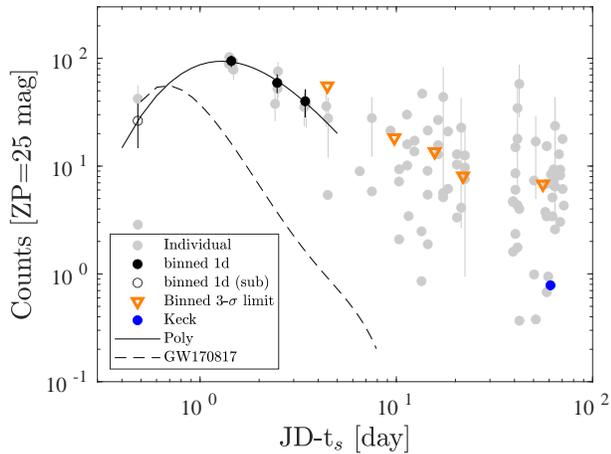}}
\caption{$g$-band light curve of AT\,2018lqh. Time is measured relative to $t_{\rm s}=2458310.348$.
Gray points represent single ZTF photometric measurements.
Black-filled circles show one day binned photometric points,
while the orange points represent bins with an adaptive size (a few days; see text).
Triangles represent $3-\sigma$ upper limits, while the Keck measurement
is shown as a blue circle.
The empty black circle represents marginal detection ($2.2\sigma$).
The solid line shows a polynomial fit to the measurements,
while the dashed line is the AT\,2017gfo (GW\,170817) afterglow
$g$-band light curve adopted from \cite{Waxman+2018_GW170817_LC},
corrected to the distance of AT\,2018lqh.
\label{fig:ZTF18gpl_g}}
\end{figure}

\begin{figure}
\centerline{\includegraphics[width=8cm]{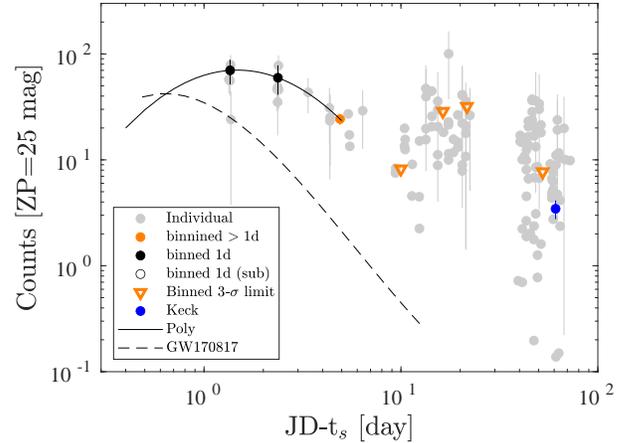}}
\caption{Same as Figure~\ref{fig:ZTF18gpl_g}, but for the $r$ band.
\label{fig:ZTF18gpl_r}}
\end{figure}

The ZTF image reduction is described in \cite{Masci+2019_ZTF_Pipeline}.
Photometry is carried out using a point-spread function fitting over
the subtracted images obtained using the 
\cite{Zackay+2016_ZOGY_ImageSubtraction} algorithm.
Table~\ref{tab:phot} lists all the photometric observations of AT\,2018lqh,
and its $g$- and $r$-band light curves are presented in Figures~\ref{fig:ZTF18gpl_gr}--\ref{fig:ZTF18gpl_r}.

Inspection of the pre-discovery ZTF data indicates that the transient
is detected at the 2.2\,$\sigma$ level (in binned $g$-band data) about one day before the first
automated detection
(JD 2458310.8).
Time is measured relative to $t_{\rm s}=2458310.348$\,day,
which is roughly half a day prior to the
first 2.2\,$\sigma$ marginal detection (empty black circle in Figure~\ref{fig:ZTF18gpl_g}).
Judging by the non detection near $t_{\rm s}$, it is likely that the time of zero flux
is roughly equal or larger than $t_{\rm s}$, but earlier times of zero flux cannot be ruled out.

\begin{figure}
\centerline{\includegraphics[width=8cm]{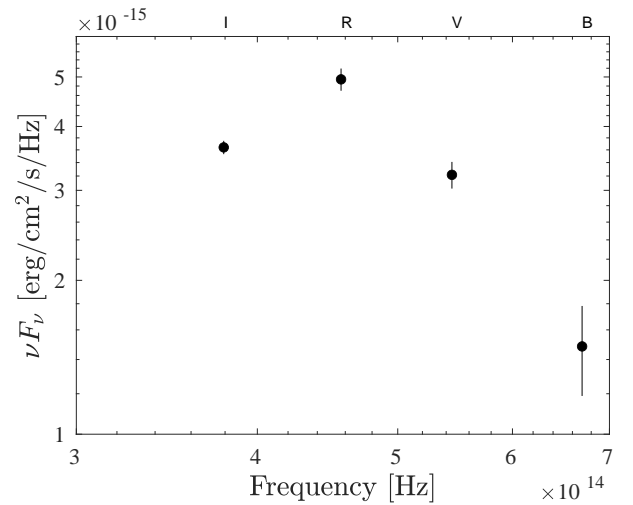}}
\caption{$\nu F_{\nu}$ vs. frequency from the the late time $BVRI$ Keck photometry. 
\label{fig:LateTimePhotSpec}}
\end{figure}

In addition to the ZTF photometry,
we obtained Keck/LRIS $BVRI$ imaging of the source on 2018 Sep 10 and 2019 Mar 7
(61\,days and 239\,days after $t_{\rm s}$, respectively).  The data were reduced and photometrically calibrated using \texttt{lpipe} (\citealt{Perley2019_lpipe}).  Adopting the second epoch images as references, we aligned and subtracted the first epoch images using \texttt{ISIS} (\citealt{Alard+1998_subtraction}; \citealt{Alard2000_subtraction_varyingKernel}).  The difference images show a clear decrease in flux at the location of the transient between 61 and 239\,days after the estimated time of zero flux.
In the $t=61$\,day epoch, we measured aperture (AB) magnitudes of $I=23.98\pm0.03$, $R=23.86\pm0.05$, $V=24.53\pm0.06$, and $B=25.61\pm0.20$.
The magnitudes were calculated in the Vega system and converted to the AB system assuming
a black-body spectrum with a temperature of 4500\,K.
The flux vs. frequency, as measured from
the Keck photometry at 61\,days since $t_{\rm s}$,
is shown in Figure~\ref{fig:LateTimePhotSpec}.
The late time emission is not consistent
with a flat $\nu F_{\nu}$, and it deviates from
a black-body spectrum.
The result regarding the rough shape of the late time spectral energy distribution is not very sensitive to the extinction.

\begin{deluxetable*}{llrllll}
\tablecolumns{7}
\tablewidth{0pt}
\tablecaption{ZTF Photometric measurements of AT\,2018lqh}
\tablehead{
\colhead{JD-2450000}    &
\colhead{Band}          &
\colhead{Counts}        &
\colhead{Counts error}  &
\colhead{ZP}           &
\colhead{Mag}           &
\colhead{$S/N$}        \\
\colhead{(day)}       &
\colhead{}            &
\colhead{(cnt)}       &
\colhead{(cnt)}     &
\colhead{(mag)}     &
\colhead{(mag)}     &
\colhead{}
}
\startdata
 8222.9298 & 2 &   $-28.83$ &    30.57 &  26.275  & 27.39  &  0.94\\
 8288.7745 & 2 &   $ 11.36$ &    26.38 &  26.275  & 23.56  &  0.43\\
 8288.7801 & 2 &   $ 44.10$ &    22.10 &  26.275  & 22.16  &  2.00\\
 8288.8010 & 2 &   $-10.27$ &    24.64 &  26.275  & 26.35  &  0.42\\
 8288.8024 & 2 &   $ 18.96$ &    21.73 &  26.275  & 23.05  &  0.87
\enddata
\tablecomments{Image-subtraction-based light curve of AT\,2018lqh.
Band 1 and 2 are $g$ and $r$, respectively.
First five lines are given. 
The full table is available in the electronic version of the paper.}
\label{tab:phot}
\end{deluxetable*}
Figure~\ref{fig:ZTF18gpl_gr} presents the $g$- and $r$-band flux residual light curve of
AT\,2018lqh prior and post $t_{\rm s}$.
Figures~\ref{fig:ZTF18gpl_g} and \ref{fig:ZTF18gpl_r} show the $g$- and $r$-band binned light curves, respectively.
The gray points represent single observations (flux residual measurements),
the black-filled points show nightly bins (in case of $>3$-$\sigma$ detection),
while the orange points/triangles are of several night bins, with bins calculated between times of 
3.5, 5.5, 11.5, 20.5, 30.5, and 100\,days after $t_{\rm s}$.
The blue points show the late time Keck observations interpolated to the $g$ and $r$ band.

\begin{figure}
\centerline{\includegraphics[width=8cm]{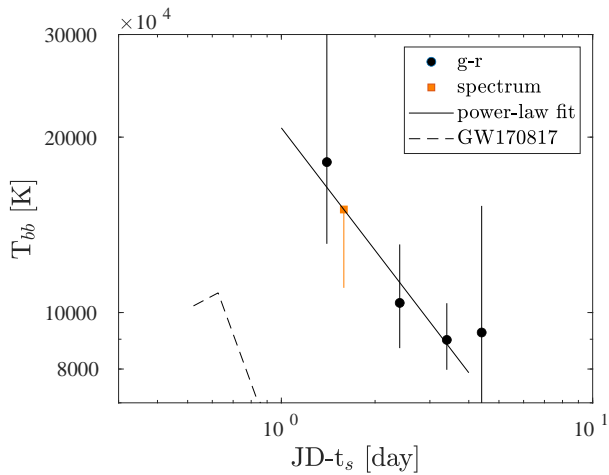}}
\caption{The effective temperature as measured from the $g$- and $r$-band data points only. Given that this is based on only two wavelengths, this estimate is highly uncertain.
However, the consistency of the photometric- and spectroscopic-derived temperatures suggests that our estimate is reasonable.
The orange square denotes the temperature measured from the spectrum (after correcting for Galactic extinction).
This data point is in agreement with the photometric-based points.
Assuming the power-law time is measured relative to $t_{\rm s}$, the effective temperature as a function of time is best fitted as a power-law with power-law index
of $\approx-0.7\pm0.2$ ($\chi^{2}/dof=0.1/2$).
The dashed line shows the estimated effective temperature as a function of the time of AT\,2017gfo (based on multi-color data; \citealt{Waxman+2018_GW170817_LC}).
\label{fig:ZTF18gpl_T}}
\end{figure}

\begin{figure}
\centerline{\includegraphics[width=8cm]{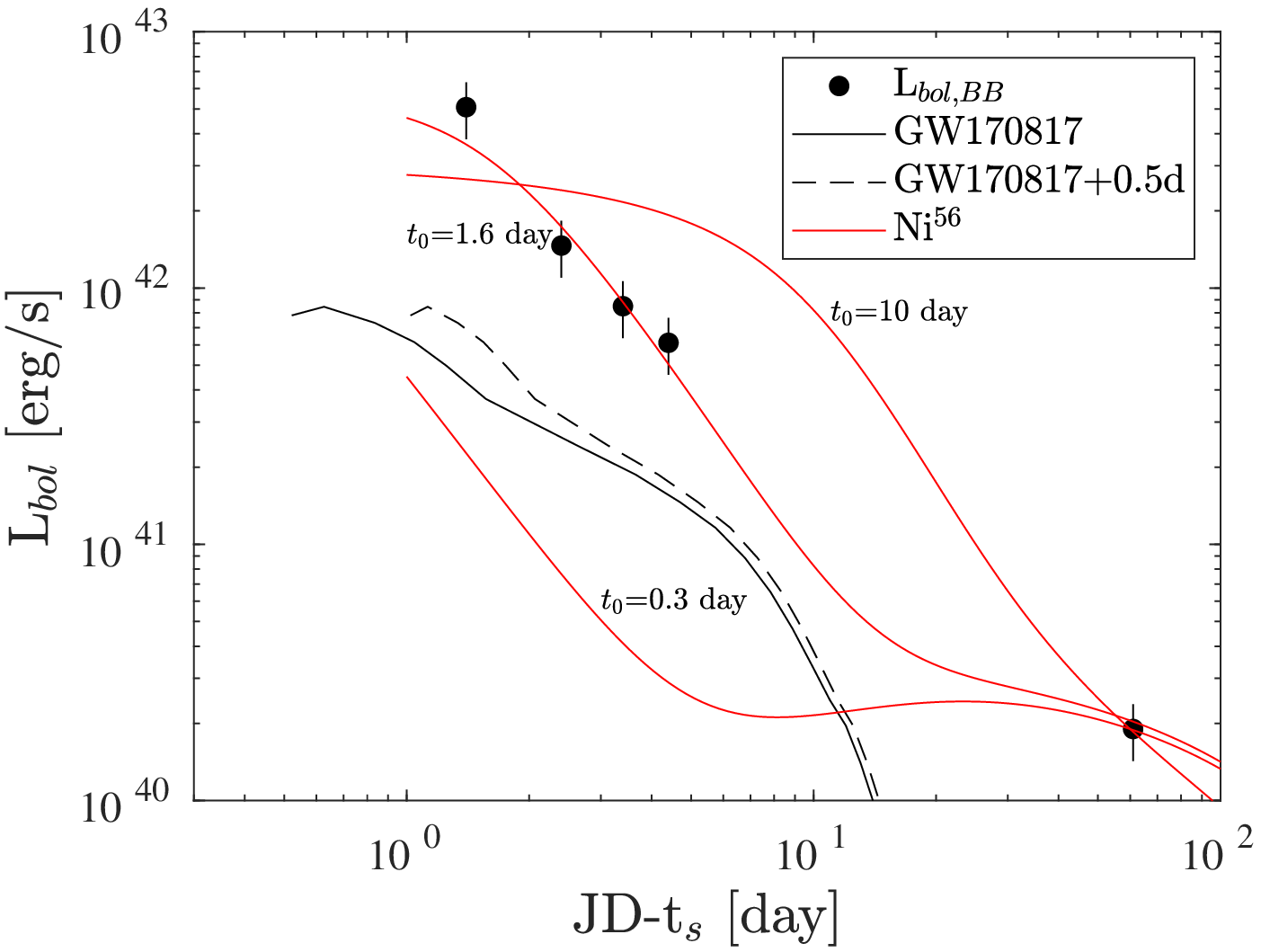}}
\caption{The approximated bolometric light curve of AT\,2018lqh as derived from integrating
the black-body fits
(see Figure~\ref{fig:ZTF18gpl_T}),
while the last (Keck) point is estimated via trapezoidal integration.
The red solid line shows the total
deposited energy by 
$^{56}$Ni, with $t_{0}$ of 0.3, 1.6 and 10\,days (\citealt{Sharon+Kushnir2020_gammaRaysDeposition_CoreCollapseSupernovae}).
The $t_{0}$ of $0.3$ and $10$\,days were normalized such that they pass through the late time point, while the $t_{0}=1.6$\,days is the best fit line (see \S\ref{sec:radioactive}).
The $^{56}$Ni mass required for a $t_{0}$ of 0.3, 1.6 and 10\,days is 0.07, 0.07 and 0.04\,M$_{\odot}$, respectively.
The black solid and dashed lines show the bolometric light curve of AT\,2017gfo (GW\,170817) assuming different
explosion times of $t_{\rm s}$ (solid line) and $t_{\rm s}+0.5$\,day (dashed line).
The bolometric light curve of AT\,2017gfo is adopted from \cite{Waxman+2018_GW170817_LC}.
\label{fig:ZTF18gpl_bolLC}}
\end{figure}

In order to estimate the bolometric light curve,
we fitted a black-body curve to the ZTF $g$-
and $r$-band data.
The best fitted temperature as a function of time is shown in Figure~\ref{fig:ZTF18gpl_T}.
The black-body radius evolution is not well constrained and is likely slower than $R\propto t^{1/2}$.
The bolometric luminosity at $t=61$\,days (the Keck epoch) is estimated by the trapezoidal
integration of the $BVRI$ data,
and hence should be regarded as a lower limit.
The errors in the bolometric luminosity were set to 25\% of the luminosity.
These errors were chosen such that our best-fit Ni$^{56}$-radioactive-energy-deposition resulted in $\chi^{2}/dof\approx1$ (see \S\ref{sec:radioactive}).
The bolometric luminosity is presented in Figure~\ref{fig:ZTF18gpl_bolLC}
and listed in Table~\ref{tab:bol}.

In Figures~\ref{fig:ZTF18gpl_g}, \ref{fig:ZTF18gpl_r}, and \ref{fig:ZTF18gpl_bolLC},
the $g$- and $r$-bands, and the bolometric light curves,
respectively,
of the GW\,170817 optical counterpart are shown
as solid black lines.
It is clear that AT\,2018lqh is brighter and has a slower time evolution compared with AT\,2017gfo.
\begin{deluxetable}{lllc}
\tablecolumns{4}
\tablewidth{0pt}
\tablecaption{Estimated bolometric lumionosity of AT\,2018lqh}
\tablehead{
\colhead{$JD-t_{\rm s}$}  &
\colhead{Luminosity}           &
\colhead{Temperature} &
\colhead{Source}     \\
\colhead{(day)}           &
\colhead{(erg\,s$^{-1}$)} &
\colhead{(K)}             &
\colhead{}
}
\startdata
1.4 & $5.1\times10^{42}$  & $18000_{-5000}^{+16000}$ & ZTF\\
2.4 & $1.5\times10^{42}$  & $10400_{-1700}^{+2700}$ & ZTF\\
3.4 & $8.5\times10^{41}$  & $9000_{-1000}^{+1400}$ & ZTF\\
4.4 & $6.1\times10^{41}$  & $9200_{-2500}^{+6000}$ & ZTF\\
61  & $1.9\times10^{40}$  &                        & Keck
\enddata
\tablecomments{Estimated bolometric luminosity of AT\,2018lqh (see text for details).
The last (Keck) data point is based on trapezoidal integration and hence should be regarded as a lower limit. Also given are the best-fit black-body temperatures.}
\label{tab:bol}
\end{deluxetable}

\subsection{Spectroscopy}
\label{sec:spec}

\begin{figure}
\centerline{\includegraphics[width=8cm]{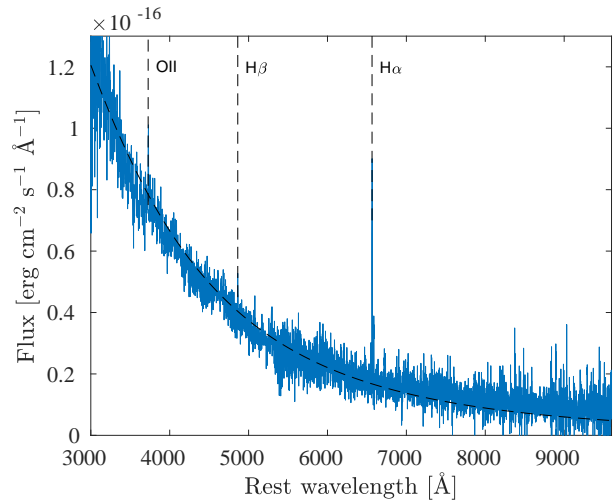}}
\caption{The spectrum of AT\,2018lqh (blue line). The best fit black body
fit ($T=15200$\,K) is shown as a black dashed line, while the H$\alpha$,
H$\beta$ and \ion{O}{2} lines are indicated by vertical lines.
\label{fig:spec}}
\end{figure}

A few hours after the first detection,
we obtained a spectrum
(at JD 2458311.9389; 1.6 days after $t_{\rm s}$)
of the event using the Low Resolution Imaging Spectrograph (LRIS; \citealt{Oke+1995_LRIS})
mounted on the Keck-I 10-m telescope.
The 560-nm dichroic was used with the 400/3400 grism
and 400/8500 grating, and a $1''$ slit.
The spectrum, presented in Figure~\ref{fig:spec},
shows a blue continuum, with narrow Balmer and \ion{O}{2} emission lines.
The $H\alpha$ flux in the SDSS spectrum is about $8\times10^{-16}$\,erg\,cm$^{-2}$\,s$^{-1}$,
while the $H\alpha$ flux in the transient spectrum is about $4\times10^{-16}$\,erg\,cm$^{-2}$\,s$^{-1}$
(after calibrating the spectrum to match the ZTF photometry).
Given that the transient is about $4''$ off the host galaxy center,
while SDSS spectra are typically centered on galaxies and have 1--$1.5''$ radius fibers,
we cannot rule out that
some of the emission lines originate from the transient.

In order to constrain any additional extinction
due to the host galaxy, we searched for a
\ion{Na}{1}D absorption line in the transient
spectrum. By measuring the equivalent width
of the wavelength region in which the
doublet is expected, we were able to put an upper limit of about 1\,\AA~on the equivalent width of the line.
Using the rough relation of \cite{Poznanski+2012_SodiumAbsorption_Dust_Extinction_Relation}, this upper limit is translated to an upper limit
on $E_{\rm B-V}$ of about $0.5$\,magnitude.

\begin{figure}
\centerline{\includegraphics[width=8cm]{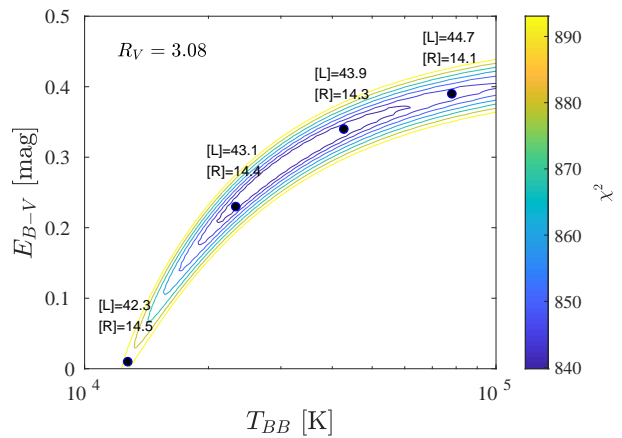}}
\caption{The best fit black-body
temperature as a function of the extinction $E_{B-V}$.
The fit was performed on the optical spectrum taken at $t_{\rm s}+1.6$\,days.
We assumed a selective extinction of $R_{V}=3.08$.
The contours show the 1 to 7-$\sigma$ confidence levels.
Along the best fit region, we marked a few points (black circles) with the $\log_{10}$ of the best fit radius
(cm; bottom number) and best fit luminosity (erg\,s$^{-1}$; upper number).
The confidence levels are calculated assuming $\chi^{2}$ statistics with three degrees of freedom.
The colorbar shows the $\chi^{2}$ value,
where the minimum $\chi^{2}$ is 846.
We note that the spectrum is measured at about 4000
wavelengths, however, the number of independent
wavelength is effectively a few times lower.
\label{fig:Fit_TRE_Rv308}}
\end{figure}

\begin{figure}
\centerline{\includegraphics[width=8cm]{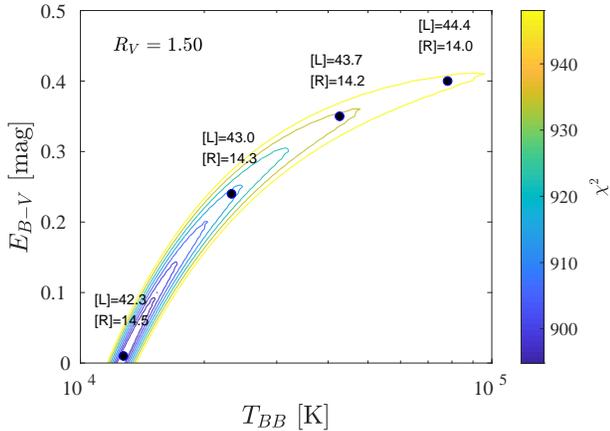}}
\caption{Same as Figure~\ref{fig:Fit_TRE_Rv308}, but for $R_{V}=1.5$.
\label{fig:Fit_TRE_Rv15}}
\end{figure}

We transformed the spectrum to the rest frame by
multiplying the wavelength by $(1+z)$ and the specific flux (per wavelength) by $(1+z)$.
We removed the region containing the H$\alpha$ line
from the spectrum and fitted a black body.
The flux density uncertainty in each wavelength bin, was estimated by applying
a running standard deviation filter to the spectrum, with a block size of 20 points.
In addition, we multiplied the errors by 1.3, to account for possible physical deviations from a black body.
The last step entails mainly increasing our uncertainties on the fitted parameters.
In Figures~\ref{fig:Fit_TRE_Rv308}--\ref{fig:Fit_TRE_Rv15}, we present the best fit black-body
temperature as a function of the extinction $E_{B-V}$ for
the selective extinction of $R_{V}=3.08$
(\citealt{Cardelli+1989_Extinction}) and $R_{V}=1.5$, respectively.  
The contours show the 1 to 7-$\sigma$ confidence levels.
Along the best fit region, we marked a few points (black circles) with the $\log_{10}$ of the best fit radius
(bottom number) and best fit luminosity (upper number).
At low extinctions,
our fit prefers $T=15,000$\,K, 
$R\approx3\times10^{14}$\,cm,
and $L=2\times10^{42}$\,erg\,s$^{-1}$.
As the plots suggest,
assuming $R_{V}\approx3$, our fit prefers a high extinction of $E_{B-V}\approx0.3$\, mag.
Such a high extinction requires
roughly an order-of-magnitude higher bolometric luminosity and a somewhat smaller radius.
However, 
$R_{V}\approx1.5$ prefers low-extinction values.
The effect of a possible extinction on the interpretation is further explored in \S\ref{sec:disc}.


The black-body fit and the derived temperatures are
on one hand critical to our interpretation, and on the other hand
may suffer from large uncertainties (e.g., due to the lack
of measurements at shorter wavelengths).
Furthermore, the black-body fit to the spectrum shows some systematic
mismatch at the red-end of the spectrum, that can be attributed
to incorrect removal of atmospheric extinction, or to deviations
from a pure black-body.
A somewhat reassuring fact is that the black-body temperature estimated
from the Keck spectrum is consistent with the temperature obtained
from the two-band photometry (see Fig.~\ref{fig:ZTF18gpl_T}).
Nevertheless, under the assumption that the spectrum, to first order,
is described by a black-body, figures 8 and 9 show the range of possible
temperatures
after renormalizing the errors such that $\chi^{2}/$dof$=1$ at
the best fit value.
This means that, to first order, deviations from a pure black-body spectrum
are taken into account in our uncertainties shown in these plots.

The photometric and spectroscopic data is available via WISeREP\footnote{www.wiserep.org} (\citealt{Yaron+GalYam2012_WiseREP}).

\subsection{Radio observations}

We observed the field of AT\,2018lqh using the AMI large array
(AMI-LA; \cite{} Zwart et al. 2008; Hickish et al. 2018) radio telescope at a central frequency of 15.5 GHz (5 GHz bandwidth) with a synthesized beam size of
$\approx 30''$. Overall, we carried out six AMI-LA observations starting on 2018 December 19, about 160 days after optical detection, and up to 2019 February 15.
The Initial data reduction, flagging and calibration of the phase and flux was carried out using {\tt reduce\_dc}, a customized AMI data reduction software package.
Phase calibration was conducted using short interleaved observations of J1613$+$3412, while for absolute flux calibration, we used 3C286.
Additional flagging and imaging were preformed using CASA.
Our AMI-LA observations revealed a radio source at the phase center with a flux level variation in the range of $130 –- 200\mu$Jy between the different observing epochs.

 Following the results from the AMI-LA observation, we undertook a radio observation of AT\,2018lqh at a higher angular resolution using the Jansky Very Large Array (JVLA)
 under the director discretionary time program 19A-452.
 The VLA observations were carried out on 2019 February 14 (218 days after $t_{\rm s}$), in both $5$\,GHz (C-band) and $13$\,GHz (K$_{\rm u}$-band). The data were reduced and imaged using standard CASA routines,
 where J1613$+$3412 was used as a phase calibrator and 3C286 as a bandpass and flux calibrator.
 The K$_{\rm u}$-band observation resulted in a null-detection with a $3\sigma$ limit of $36\mu$Jy. A null-detection was also the result of the C-band observations with a $3\sigma$ limit of $21\mu$Jy, although a weak ($\sim 35\mu$Jy) point source was detected approximately $3''$ away from the optical position of AT\,2018lqh.
 The C-band limit translates to $L_{\nu}\ltorder1.5\times10^{27}$\,erg\,s$^{-1}$\,Hz$^{-1}$,
 or $\nu L_{\nu}\ltorder8.8\times10^{36}$\,erg\,s$^{-1}$.
 Comparing the results of the AMI-LA observations with the VLA observations, we conclude that the origin of the radio emission we detected with AMI-LA is probably diffuse emission from the host galaxy (resolved out by the VLA), and not from AT\,2018lqh.

\subsection{X-Ray observations}

On 2019 Feb 11 (215\,days after $t_{\rm s}$), we obtained a 9574\,s integration
of the source using {\it Swift}-XRT (\citealt{Gehrels+2004_Swift}).
We used the Poisson-noise matched filter (\citealt{Ofek+2018_PoissonNoiseMF}) to search for
point sources (with the XRT point spread function) at the location
of the transient.
We marginally detected an X-ray source at the transient location,
with a false-alarm probability of 0.002.
The X-ray source has three photons in a 10-arcsecond radius aperture,
in the 0.2--10\,keV range.
Assuming $z=0.05446$, $n_{\rm H}=1.48\times10^{20}$\,cm$^{-2}$,
and a power-law spectrum with a photon index $\Gamma=1.6$,
we get\footnote{Calculated using WebPIMMS: https://heasarc.gsfc.nasa.gov/cgi-bin/Tools/w3pimms/w3pimms.pl} an unabsorbed flux of $(1.4_{-0.7}^{+1.4})\times10^{-14}$\,erg\,s\,cm$^{-2}$.
This corresponds to an X-ray luminosity of $L_{0.2-10 \rm keV}=9.8\times10^{40}$\,erg\,s$^{-1}$.
This X-ray luminosity is consistent with the X-ray luminosity
reported for Type IIn SNe, but in time scales of weeks after the explosion (e.g., \citealt{Ofek+2013_IIn_X}).
However, given the non-coincidence of the radio source
with the transient, it is possible that the X-ray source is associated with the
diffused radio emission
and that it is unrelated to the supernova.


\section{The Nature of the Short-Duration Transient}
\label{sec:disc}

The early, day-time scale, emission from supernova explosions is well explained in many cases as the result of the escape of radiation from the expanding shock-heated outer layers of the exploding star ({\it envelope cooling}), or of the escape of photons from a radiation-mediated shock driven by these layers into the CSM ({\it CSM breakout}). We show here that the high luminosity and rapid evolution of ZTF\,18lqh implies that the radiation emitted in this event is unlikely solely dominated by envelope cooling nor CSM breakout.
However, we cannot rule out that the early-time emission is dominated by the CSM interaction,
while the late-time emission is due to shock cooling. Alternatively, the unique properties of ZTF\,18lqh suggest that its radiation was powered by radioactive decay in a rapidly expanding low-mass shell of a highly radioactive low-opacity material.

\subsection{The envelope-cooling scenario}
\label{sec:cooling}

Let us examine first the envelope-cooling scenario. Consider a shell of mass $m$ that is expanding at velocity $v$, with a thickness $\Delta r$ dominated by the velocity spread $\Delta v$ across the shell, $\Delta r=\Delta{v} t$. As long as the shell's optical depth $\tau$ is large, such that the radiation (diffusion) escape time $\tau\Delta r/c$ is large compared to the expansion time $t$, the luminosity produced by the escaping radiation is roughly constant: The energy $E$ carried by the radiation decreases as $E\propto 1/r$ due to adiabatic expansion, the optical depth decreases as $\tau\propto 1/r^2$, and $L\approx E/(\tau\Delta r/c)\propto r^0$. The luminosity of ZTF\,18lqh begins to decline at $t\lesssim 1.5$~d, and drops between 1 and 5~d with a characteristic time scale of 1~d ($L$ drops by a factor of 6(8) over 2(3)~d, corresponding to an exponential decay with a 1.1(1.4)~d time scale). This suggests that $\tau\Delta r/c\sim t$ at $t=1.5$\,days.
Using $\tau\approx\kappa m/(4\pi r^2)$, where $\kappa$ is the opacity, we have an estimate for the mass of the envelope $m$:
\begin{equation}
    \label{eq:m}
    m\approx \frac{r}{\Delta r}\frac{4\pi rct}{\kappa}\cong
    0.02\, \frac{r}{\Delta r}\frac{r_{14.5}}{\kappa_{0.34}}\frac{t}{1.5~\rm d}M_\odot.
\end{equation}
Here, $r\equiv10^{14.5}r_{14.5}$\,cm, $\kappa\equiv0.34\kappa_{0.34}\,{\rm cm^2}$\,g$^{-1}$.
Next, the shell's velocity is:
\begin{equation}
    \label{eq:v}
    v\approx \frac{r}{t}\cong0.08\, r_{14.5}\left(\frac{t}{1.5~\rm d}\right)^{-1}\,c,
\end{equation}
where we have used the radius obtained from the black-body fit (\S\ref{sec:phot}--\ref{sec:spec}).
For the case of high extinction (i.e., $E_{B-V}\approx0.4$\,mag; Fig.~\ref{fig:Fit_TRE_Rv308}), the radius is $\approx1.3\times10^{14}$\,cm,
and the velocity $\approx0.03$c (more typical of SN velocities).
Since we use $t_{\rm s}$ to estimate
properties like the ejecta velocity and mass, we need to discuss
some possible uncertainties regarding this reference time.
Equations~\ref{eq:m} and \ref{eq:v} can be used as a definition of $t_{\rm s}$.
If $t_{\rm s}$ is half a day later (which is the upper limit on $t_{\rm s}$),
then the estimated velocity will be 30\% higher.
On the other hand, in order for the velocity to be closer to
the typical values observed in SNe, $t_{\rm s}$ should be about 4\,days,
which, in turn, will require a dark period in the transient evolution.
Furthermore, in such case, the estimated 
ejected mass will only be a few times higher than the nominal value in
Equation~\ref{eq:v}.

Using this velocity,
we also have an estimate of the shell's kinetic energy:
\begin{equation}
    E_k\approx \frac{r}{\Delta r}\frac{4\pi r^3c}{2\kappa t}
    \cong 1.2\times10^{50} \frac{r}{\Delta r}\frac{r^3_{14.5}}{\kappa_{0.34}}\left(\frac{t}{1.5~{\rm d}}\right)^{-1}
    \rm erg.
    \label{eq:kinetic_energy}
\end{equation}
The value of $r$ is normalized in the above equations to the inferred value at $t=1.5$\,d, and the value of $\kappa$ is normalized to the opacity of a 70:30
Hydrogen-to-Helium mix by mass.

Assuming negligible extinction, the radiated energy, $E_r\approx 7\times10^{47}$\,erg, over the first days is $\approx 100$ times smaller than the shell's kinetic energy. Since the shock that accelerated the shell generated similar kinetic and thermal energies, and since the thermal energy drops like $1/r$, the initial radius of the shell (the radius at which the shock passed it) should have been $\approx (E_r/E_k)r\approx 10^{12}$\,cm.
Thus, the envelope-cooling explanation requires $\approx10^{50}$\,erg to be deposited in the outer $\approx0.01$\,M$_\odot$ shell of a $\approx10^{12}$\,cm star, accelerating it to $\approx0.08$\,c.
Depositing such a large amount of energy in the outer 0.01~$M_\odot$ shell of a supergiant star, which would accelerate it to $0.08$c, is challenging.
It would require orders of magnitude larger energy to be deposited in the inner, more massive stellar shells.
On the other hand, if the extinction is high ($E_{B-V}\approx0.4$\,mag; see Figure~\ref{fig:Fit_TRE_Rv308}), then the radiated energy is $E_{r}\sim2\times10^{50}$\,erg (luminosity multiplied by about four days).
In addition, in this case, the radius is smaller by a factor of 2.5, and, therefore, the kinetic energy is smaller by an order of magnitude (Eq.~\ref{eq:kinetic_energy}), and $E_{r}>E_{k}$.
For intermediate values of the extinction, $E_r\sim E_k$,
this requires a $\approx10^{14}$\,cm star.
Both the large star requirement and the efficient release of the equivalent of the kinetic energy within a short time
frame are challenging and seem unlikely.

\subsection{The CSM breakout scenario}

Let us consider next the CSM breakout scenario
(e.g., \citealt{Ofek+2010_PTF09uj_windBreakout};
\citealt{Katz+2011_Xrays_gamma_neutrinos_FromCollisionlessShocks_SupernovaWindBreakouts}).
In this case, photons escape the radiation-mediated shock when the optical depth of the CSM lying ahead of the shock is comparable to $c/v$, where $v$ is the shock velocity (\citealt{Weaver1976_StructureOfSupernovaeShocks}). The mass and kinetic energy of the shocked CSM layer are thus approximately given by equations \ref{eq:m}-\ref{eq:kinetic_energy}, with $\Delta r/r$ set to 1.
In the CSM-breakout scenario, the radiated energy is expected to be similar to the kinetic energy of the shocked CSM, $E_r\approx E_k$, since the post-shock kinetic and thermal energies are similar. 
For the zero extinction case,
this is inconsistent with the observed $E_r/E_k\sim10^{-2}$.
However, equating the kinetic energy estimate (Equation~\ref{eq:kinetic_energy})
with the best fit luminosity, as a function of $E_{B-V}$, in Figure~\ref{fig:Fit_TRE_Rv308} (multiplied by four days),
lead to the suggestion that a $E_{B-V}\approx0.3$\,mag
(with $T\approx4\times10^{4}$\,K and $E_{k}\approx3\times10^{49}$\,erg) is a viable solution.
Such a scenario will still require a low amount of CSM mass, $\sim10^{-2}$\,M$_{\odot}$.

Assuming that the emission in the first few days is produced by a CSM breakout implies that the shock driven at later times by the ejecta into the CSM at larger radii is collisionless \citep{Katz+2011_Xrays_gamma_neutrinos_FromCollisionlessShocks_SupernovaWindBreakouts}. The emission observed at 60\,days may be produced in this case both by radiation escaping from a massive shock-heated ejecta and by synchrotron emission produced by the collisionless shock. 

Let us consider first the cooling massive ejecta option. As long as the photon diffusion time through the ejecta is larger than the expansion time $t$, the (constant) luminosity $L$ is approximately given by:
\begin{eqnarray}
    L & \approx & 2\pi R_{0} v^{2} \frac{c}{\kappa} \\
    & \cong & 1.4\times10^{40} \frac{R_0}{10^{11}\,{\rm cm}} \Big(\frac{v}{5000\,{\rm km\,s}^{-1}} \Big)^{2} \Big( \frac{\kappa}{0.34\,{\rm cm^{2}\,g}^{-1} } \Big)^{-1},
\end{eqnarray}
where $R_0$ is the stellar radius, $v$ is the expansion velocity and $\kappa$ is the opacity (this is obtained noting that $L\approx E/(\tau r/c)$ with $E$, the radiation energy stored in the ejecta, given by $E\approx E_0 (R_0/r)$ with initial energy $E_0\approx 0.5 M v^2$ and $\tau\approx \kappa M/(4\pi r^2)$). The observed luminosity, $L\approx10^{40}$\,erg\,s$^{-1}$, may be obtained for a rather compact progenitor, $R_0\approx10^{11}$\,cm with $v\approx5\times10^3$\,km\,s$^{-1}$ and $M>7M_\odot$ (to ensure $\tau>c/v$ at 60\,d).

Let us consider next the collisionless shock emission. Shock-accelerated electrons emitting synchrotron radiation in the optical band are expected to cool fast, losing most of their energy over a time much shorter than $t$. Given that shock acceleration is expected to produce an electron energy distribution with equal energy per logarithmic interval of electron Lornetz factor, $\gamma^2 dn/d\gamma=const.$, the synchrotron luminosity in the optical band is $L\approx 2\pi (\varepsilon_e/\Lambda)r^2 \rho v^3$, where $\varepsilon_e\approx0.1$ is the fraction of shock energy carried by accelerated electrons, $\Lambda\approx 2\ln(\gamma_{\max}/\gamma_{\min})\sim 10$, and $\rho$ is the CSM density. The CSM mass required to produce the observed luminosity is $M\approx 2\Lambda Lt/(v^2\varepsilon_e)\approx 0.01M_\odot$ for $v=10^4{\rm km/s}$.
However, if the late time radiation
is dominated by synchrotron emission, we expect the spectrum to be rather blue (with $\nu L_{\nu}\sim const$).
This is in contrast to the
observed late time colors of the transient (Fig.~\ref{fig:LateTimePhotSpec}).

In a CSM interaction scenario, we expect to detect narrow-to-intermediate-width emission lines (e.g., \citealt{Chevalier+Fransson1994_Emission_SN_CSM_Interaction}).
Although an H$\alpha$ emission line is present
in the transient spectrum (Fig.~\ref{fig:spec}),
this line is unresolved, and its luminosity is roughly consistent with the H$\alpha$ line luminosity in the host-galaxy spectrum, as obtained by SDSS.
A possible explanation for the lack of emission lines
was suggested by \cite{Moriya+Tominaga2012_IIn_nonSteadyMassLoss_lines}.
Specifically, they argue that if the CSM has a density profile $\rho\propto R^{-w}$
with a cutoff and where $w\ltorder 1$, the shock can go through the entire CSM only after the light curve peak,
and in this case, narrow lines from the CSM will be weak or absent.
Another possible explanation is that the temperature is high and, hence, the Balmer lines are weak.
Indeed, our high-extinction scenario requires
a high effective temperature ($T\sim5\times10^{4}$\,K;
Figure~\ref{fig:Fit_TRE_Rv308}).
However, in such cases, we expect to detect higher ionization species (e.g.,
\citealt{Chevalier+Fransson1994_Emission_SN_CSM_Interaction};
\citealt{Gal-Yam+2014_SN2013cu_FlashSpectroscopy};
\citealt{Yaron+2017_PTF13dqy_HighIonization_FlashSPectroscopy}).
We note that this scenario can explain, as discussed in the literature,
objects like AT\,2018cow (\citealt{Perley+2019_SN2018cow}) and
AT\,2020xnd (\citealt{Perley+2020_AT2020xnd_FastLuminousTransient}).
In these cases, intermediate width emission lines are indeed observed.

To summarize, we cannot rule out that the early emission from AT\,2018lqh is due to CSM breakout
(although the lack of intermidiate-width emission lines is not consistent with the naive expectation).
The explanation for the late time emission from this source can be either
shock cooling from a compact star or synchrotron emission due to CSM interaction, although, the latter is in odds with the
late time colors of the event.

\subsection{The fast radioactive shell scenario}
\label{sec:radioactive}

Consider a shell of radioactive material of mass $m$ expanding at velocity $v$, with thickness $\Delta r$ dominated by the velocity spread $\Delta v$ across the shell, $\Delta r=\Delta v t$. As in the envelope-cooling scenario, the luminosity produced by the shell would not decline as long as 
\begin{equation}
   t<\tau\Delta r/c.
   \label{eq:tdiff_l_t}
\end{equation}
The rapid decline of the luminosity at 1.5\,days, therefore, requires $\tau\Delta r/c< t$ at $t\cong1.5$~d. This differs from the cooling envelope case, where the time scale for the luminosity decline is determined by the escape time of the photons and hence $\tau\Delta r/c\approx t$ is required. In the radioactive-shell scenario, the decline of the luminosity may be determined by the decline in the radioactive energy deposition rate. Hence, the decline rate sets only an upper limit to the photon escape (diffusion) time. Equation~(\ref{eq:m}) is thus modified to:
\begin{equation}
    \label{eq:mrad}
    m< \frac{r}{\Delta r}\frac{4\pi rct}{\kappa}\cong
    0.08\, \frac{r}{\Delta r}\frac{r_{14.5}}{\kappa_{-1}}\frac{t}{1.5\,\rm d}M_\odot.
\end{equation}
Here, $\kappa\equiv10^{-1}\kappa_{-1}\,{\rm cm^2}$\,g$^{-1}$, normalizing the opacity to a value that is more appropriate for partially ionized Nickel. 

For $\tau\Delta r/c< t$, photons escape the shell over a time $<t$ and the luminosity reflects the radioactive energy deposition. The radiated energy, $E_r\approx 7\times10^{47}$\,erg, over the first days requires a total energy deposition per nucleus of:
\begin{equation}
    \label{eq:Erad}
    E_{A,\rm dep}\approx\frac{E_r}{m/(A m_{\rm p})}>0.2 \frac{A}{50} 
    \frac{\Delta r}{r}\frac{\kappa_{-1}}{r_{14.5}}\frac{1.5\,\rm d}{t} \rm MeV,
\end{equation}
where $A$ is the nucleon number
and $m_{\rm p}$ is the proton mass.

Before considering the implications of this result, we should take into account the fact that not all the radioactive energy is necessarily deposited in the plasma. For $\beta$-decay, the energy is released in the form of electrons, positrons and photons with $\sim1$~MeV energy. The gamma-rays lose energy at the lowest rate, and hence are the first to escape the plasma as its column density decreases. Taking into account that $\sim 1$~MeV photons lose their energy in $\sim3$ scatterings, the fraction of photon energy deposited may be estimated as
$f_\gamma\approx \min[1, (1/3)\kappa_\gamma\rho\Delta r] \approx \min[1,  (1/3)(\kappa_\gamma/\kappa)\tau]$, where $\kappa_\gamma\approx 0.03$\,cm$^{2}$\,g$^{-1}$ (see \citealt{Longair2011_HighEnergyAstrophyics_Book}).
Thus, for $\tau\ltorder 3$ and using Equation~\ref{eq:tdiff_l_t},
\begin{equation}
    \label{eq:gamma}
    f_\gamma\approx\frac{1}{3}\frac{\kappa_\gamma}{\kappa}\tau<
    \frac{1}{3}\frac{\kappa_\gamma}{\kappa}\frac{r}{\Delta r}\frac{ct}{r}\approx
    1 \frac{r}{\Delta r}\frac{1}{r_{14.5}\kappa_{-1}}\frac{t}{1.5~\rm d}.
\end{equation}
Equation~\eqref{eq:gamma} implies that a large fraction of the $\gamma$-ray energy may be deposited in the plasma at $t\sim1.5$\,d, provided that $\kappa$ is not much larger than $0.1$\,cm$^2$\,g$^{-1}$. At a later time, $f_\gamma$ drops below unity as $1/r^2\cong 1/t^2$, and most of the $\gamma$-ray energy escapes.
The small inferred value of the opacity also implies that $Y_{\rm e}$ (i.e., the number density of electrons over the baryons number density)
cannot be much smaller than $1/2$.

Equation~\eqref{eq:Erad} implies that the bolometric luminosity during the first few days may be accounted for by the $\beta$-decay of radioactive elements with a life time of the order of days, provided that $m$ is not far below its upper bound, given by Equation~\eqref{eq:mrad}, and $\kappa$ is not much larger than $0.1$\,cm$^2$\,g$^{-1}$. Much smaller values of $m$, or larger values of $\kappa$, would require the deposited energy per nucleous, and more so the radioactive energy released per nucleous, to be larger than typically expected for $\beta$-decay. The relatively low value of the opacity is consistent with that of partially ionized Ni.

Let us consider next the luminosity observed at $t=60$~d. Assuming that the luminosity at this time is dominated by the radioactive decay of a longer-lived isotope, the ratio of the luminosity at $60$~d to the peak luminosity near 1~d would be:
\begin{equation}
    \label{eq:Lrat}
    \frac{L_{\rm peak}}{L_{60\rm d}}\approx \frac{f_{\rm peak}Q_{\rm peak}}{f_{60\rm d}Q_{60\rm d}(e^\pm)}
    \frac{\tau_{60\rm d}}{\tau_{\rm peak}}.
\end{equation}
Here, $f_{\rm peak(60)}$, $Q_{\rm peak(60)}$ and $\tau_{\rm peak(60)}$ are the fraction of nuclei, the decay of which dominates the energy production on a one-day (60\,days) time scale, the radioactive energy released in their decay, and their decay time,
respectively.
At $t=60$\,days, gamma-rays escape the shell with little energy loss; hence, only the part of the decay energy carried by electrons and positrons, $Q_{60\rm d}(e^\pm)$, is considered. Note that for $e^\pm$ energy of $\approx 1$\,MeV,
the effective opacity due to ionization energy loss is $\kappa_e\equiv (dE_e/dX)/E_e\approx2$\,cm$^2$\,g$^{-1}$, 
where $dE_e/dX$ is the energy loss per unit column density
(grammage) traversed by the electrons, and $dX = \rho dx$. 
This implies that for $m\simeq0.08\,M_\odot$ and $v/c\simeq0.08$, the energy loss time of the $e^\pm$ is shorter than the expansion time up to $\sim80$\,d 
(e.g., \citealt{Waxman+2018_GW170817_LC};
\citealt{Waxman+2019_GW170817_LateTimeLightCurve}).

For the $^{56}$Ni-to-$^{56}$Co decay chain, with $^{56}$Ni decay dominating at 1\,day and  $^{56}$Co at 60\,day, we have $f_{\rm peak}/f_{60\rm d}\approx1$, $\tau_{60\rm d}/\tau_{\rm peak}\approx13$, and $Q_{\rm peak}/Q_{60\rm d}(e^\pm)\approx 1.8\,{\rm MeV}/0.12\,{\rm MeV}=15$; hence, $L_{\rm peak}/L_{60\rm d}\approx200$, a value close to the observed ratio. The analysis presented above suggests, therefore, that the observed bolometric luminosity may have been produced by a fast, $v/c\approx 0.08$, low mass, $\approx0.07\,M_\odot$, shell dominated by radioactive $^{56}$Ni.
However, other radioactive elements cannot be ruled out without a specific examination
of their decay chain.

Figure~\ref{fig:ZTF18gpl_bolLC} demonstrates the validity of this conclusion by showing the light curves that are expected to be produced by an expanding shell of $^{56}$Ni.
The red lines shows light curves
for low-mass, fast-expanding shells, for which the photon escape time is shorter than the expansion time and the luminosity is given by the radioactive energy deposition rate.
Following \cite{Sharon+Kushnir2020_gammaRaysDeposition_CoreCollapseSupernovae},
the deposition rate is approximated by:
\begin{equation}
    Q_{\rm dep}(t) = Q_{\gamma} \frac{1}{(1 + [t/t_{0}]^{3})^{2/3} } + Q_{\rm pos},
    \label{eq:Qdep}
\end{equation}
where $Q_{\gamma}$ is the energy released in $\gamma$-rays and $Q_{\rm pos}$ is the energy released in positrons
(e.g., \citealt{Swartz+1995_GammaRayTransfer_EnergyDeposition_SN}; \citealt{Junde1999_NuclearDataSheetsForA56}).
The $(1 + [t/t_{0}]^{3})^{2/3}$ factor is an interpolation between the full $\gamma$-ray energy deposition at early times, $t\ll t_0$, and the partial deposition, with a deposited fraction of $(t_{0}/t)^{2}$, at late times, $t\gg t_0$.
The red line marked with $t_{0}=1.6$\,days is the best fit model, with
$t_{0}=1.6_{-0.3}^{+0.5}$\,days and
$M=0.07\pm0.02$\,M$_{\odot}$
($\chi^{2}/dof=2.4/3$).

Following Equation~8 in \citealt{Wygoda+2019_TypeIa_TwoLumWidthRelations}, the
value of $t_{0}$ is given by:
\begin{equation}
    t_{0} = \sqrt{\kappa_{\rm eff} t^{2} \langle \Sigma \rangle_{\rm Ni56}}.
    \label{eq:t0}
\end{equation}
Here, $\langle \Sigma \rangle_{\rm Ni56}$
is the column density of the ejecta Ni$^{56}$
and $\kappa_{\rm eff}$ is the effective opacity ($\approx0.025$\,cm$^{2}$\,g$^{-1}$ for $^{56}$Co; \citealt{Swartz+1995_GammaRayTransfer_EnergyDeposition_SN}; \citealt{Jeffery+1999_RadioActiveDecay_Deposition_t0}; \citealt{Kushnir+2020_SubChandra_Observed_t0_MassNi56_Relation}).
In a homologous expansion, 
$\langle \Sigma \rangle_{\rm Ni56} t^{2}$ is constant.
However, $\langle \Sigma \rangle_{\rm Ni56} t^{2}$ depends on the velocity and angular distributions of the ejecta, as well as on the
$^{56}$Ni fraction distribution in the ejecta.
We can perform an order-of-magnitude consistency check by scaling the opacity, $^{56}$Ni mass and velocity 
of Type Ia SNe, as well as their observed $t_{0}$ ($\approx40$\,days),
to our estimated parameters.
This suggest a $t_{0}$ of a few days is expected, and our model is consistent with the observations.
The ejecta mass derived in this scenario requires
high optical-depth at early times.
Therefore, this scenario is consistent with the
optically-thick (featureless) spectrum observed at $t_{\rm s}+1.6$\,days.
The spectrum of this event shows less features than the spectra
of AT\,2017gfo, at a similar epoch.
Since the ejecta mass and opacity of this event is presumably of the same
order of magnitude as that of AT\,2017gfo, the simplest explanation
to the differences between the spectra is the lower ejecta velocity
in AT\,2018lqh compared to AT\,2017gfo.
This lower velocity will result, in higher optical depth for AT\,2018lqh,
at the same epoch.

%
In the future, this can be used to test specific models for which the ejecta distribution
is modeled.

\section{Discussion and Summary}
\label{sec:summary}

\begin{figure}
\centerline{\includegraphics[width=8cm]{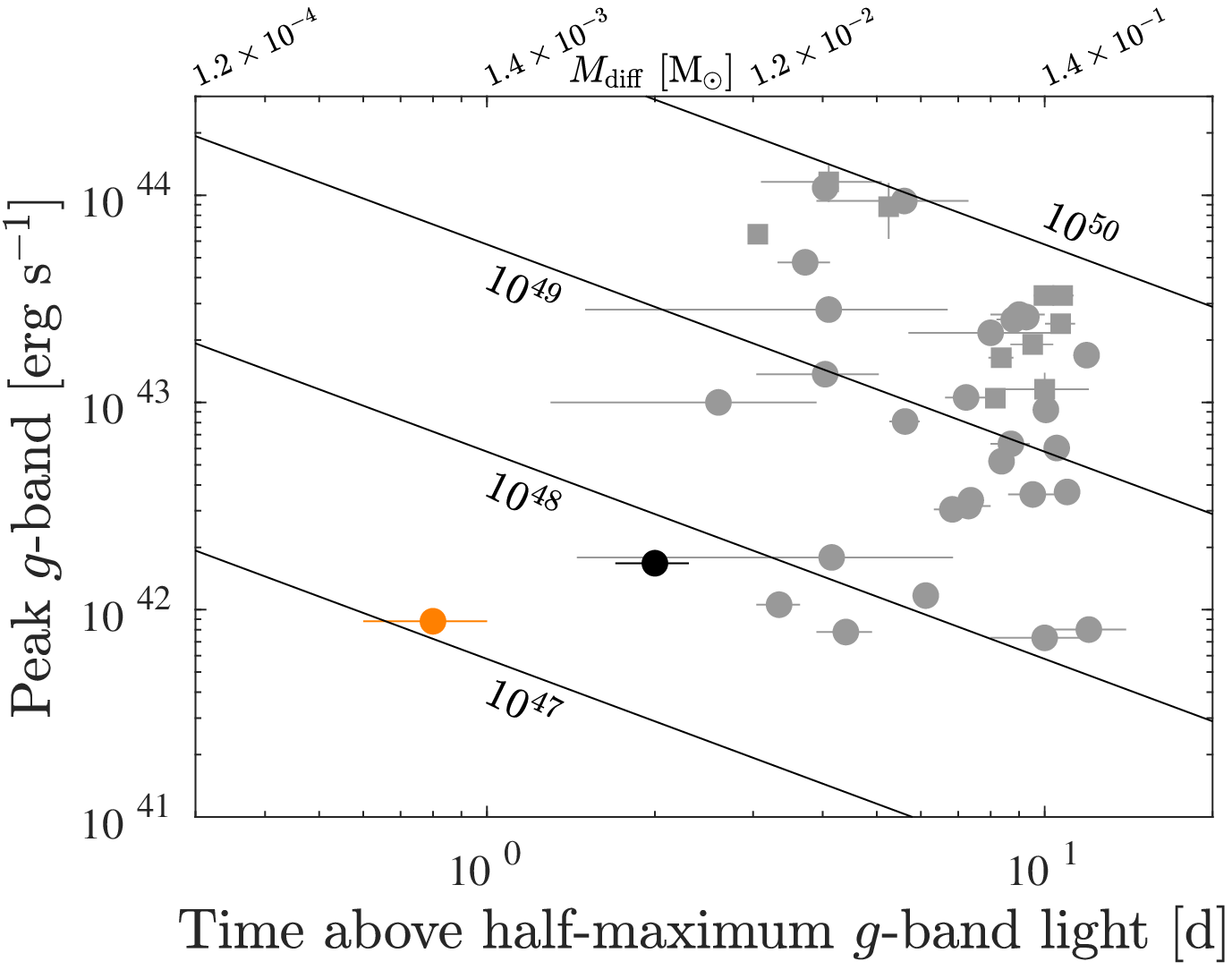}}
\caption{Time above half-maximum $g$-band light vs. peak $g$-band luminosity
for AT\,2018lqh (black circle; no extinction), AT\,2017gfo (orange circle),
and selected fast transients with a $g$-band light curve and limit, better than 3 days, on the duration (\citealt{Ofek+2010_PTF09uj_windBreakout},
\citealt{Drout+2014_Rapidly_Evolving}, \citealt{Hosseinzadeh+2017_SN_Ibn_diversity},
\citealt{Perley+2019_SN2018cow}, and the sample of \citealt{Ho+2021_ZTF_RapidlyEvolvingTransients}).
Squares depict events whose spectra show evidence of a CSM interaction (IIn/Ibn SNe),
while the rest are marked by circles.
The upper X-axis shows the estimated ejecta or CSM mass assuming that the time scale
is dominated by diffusion time, and assuming $\kappa=0.34$\,cm$^{2}$\,g$^{-1}$,
and an ejecta velocity of $10^{9}$\,cm\,s$^{-1}$ (i.e., $M_{\rm diff}\sim 4/3\pi c v t^{2}/\kappa$).
This mass roughly indicates a lower-limit on the mass in the ejecta.
The solid black lines reflect the equal total radiated $g$-band energy,
as calculated by $2L_{\rm peak}t_{1/2}$,
where $L_{\rm peak}$ is the $g$-band peak luminosity and $t_{1/2}$ is the time above half maximum $g$-band light.
The lines, from bottom to top, correspond to $10^{47}$, $10^{48}$, $10^{49}$, $10^{50}$\,erg.
\label{fig:TimeLum}}
\end{figure}

We report on the discovery of a faint, fast transient, AT\,2018lqh, using ZTF.
Based on the rough similarity of the peak
luminosity and durations, we can~not rule out
that this transient may be related to some of the fast transients
reported in \cite{Drout+2014_Rapidly_Evolving}, although it is a factor of two
faster (measured via the duration above half maximum light)
than the fastest transient reported in \cite{Drout+2014_Rapidly_Evolving}.
The observed light curve of AT\,2018lqh is brighter
and slower compared with that of the NS-NS merger event AT\,2017gfo.
Figure~\ref{fig:TimeLum} shows the peak luminosity vs. time above half-maximum $g$-band light,
with markings for the positions of AT\,2018lqh (black), AT\,2017gfo (green) and additional fast transients (\citealt{Drout+2014_Rapidly_Evolving}, \citealt{Ho+2021_ZTF_RapidlyEvolvingTransients}; gray).
Also marked on this plot are lines of equal radiated energy and the CSM mass corresponding to the diffusion time.

We discuss the possible nature of such events in the context of our spectroscopic and photometric observations.
We argue that the observations are not consistent with a shock breakout and cooling from a stellar envelope.
We suggest that there are two possible explanation for AT\,2018lqh and similar events.
The first
is a radioactive-powered transient.
This requires low-mass ejecta (a few $10^{-2}$\,M$_{\odot}$), most of which is radioactive, that has low opacity ($\kappa\ltorder0.1$\,cm$^{2}$\,g$^{-1}$) and high velocity ($v\sim 0.08c$).
The second is that the event is powered, at early times, by ejecta moving into low-mass ($10^{-2}$\,M$_{\odot}$) CSM
at a distance of about $10^{14}$--$10^{15}$\,cm from the progenitor,
while at late times, it must be powered by a different mechanism, such as shock cooling or collisionless shocks from CSM interaction.
The existence of such low-mass CSM around SN progenitors is consistent with the finding that a large fraction of Type~IIn SNe have precursors (outbursts) prior to their explosion (e.g., \citealt{Ofek+2014_IIn_precursors}; \citealt{Strotjohann+2020_ZTFPrecursorsSample_IIn}),
and that a fraction of core-collapse SNe
show evidence of a confined CSM (e.g., \citealt{Yaron+2017_PTF13dqy_HighIonization_FlashSPectroscopy}, \citealt{Bruch+2021_SN_Progenitors_ElevatedMassLoss_FlashSpectroscopy}).
However, in the context of AT\,2018lqh,
the possible absence of intermediate-width emission lines from the CSM is puzzling
(but still, this scenario can~not ruled out).

\subsection{Implications for the progenitor}

In the case that this event is powered by $^{56}$Ni radioactivity, we can roughly estimate the minimum density required of the burning material by noting that the inferred expansion velocity, $v$, should be comparable to the velocity of the shock wave that accelerated the expanding shell and heated it to a temperature $T_0\sim1{\rm\, MeV}=10^{10}$\,K, enabling burning to Nickel. Assuming that the post-shock energy density is dominated by radiation (which may be verified for the inferred density), $aT^4\sim \rho v^2$ (where $a$ is the radiation constant) and 
\begin{equation}
    \rho \gtrsim \frac{aT_0^{4}}{v^{2}} \approx 10^{7} \Big(\frac{T_0}{10^{10}\,{\rm K}}\Big)^{4} \Big(\frac{v}{0.08\,c}\Big)^{-2}\,{\rm g}\,{\rm cm}^{-3}.
\end{equation}
Such densities are presumably found in the outer layers of NS and also in the cores of white dwarfs, but the fast-shocked shell presumably lies at the outer parts of the star. A radioactive-powered scenario suggests, therefore, that a NS was involved in the explosion.
One such possible mechanism is Accretion Induced Collapse (e.g., \citealt{Dessart+2006_AIC}, \citealt{Sharon+Kushnir2019_AIC}).

For the CSM interaction scenario, an important requirement is that the star eject $\sim10^{-2}$\,M$_{\odot}$
just prior to its explosion.
Indeed, there is a variety of observational evidence that some stars eject large amounts of mass
on time scales of months to years prior to their explosion
(e.g., \citealt{Pastorello+2007_SN2006jc_precursor}; \citealt{Smith+2010_SN2009ip_UGC2773OT_precursor}; \citealt{Ofek+2013_PTF10tel}; \citealt{Ofek+2013_SN2009ip}; \citealt{Ofek+2014_IIn_precursors}; \citealt{Ofek+2014_NuSTAR}; \citealt{Gal-Yam+2014_SN2013cu_FlashSpectroscopy};
\citealt{Bruch+2021_SN_Progenitors_ElevatedMassLoss_FlashSpectroscopy}).
In addition, there are some theoretical arguments for the existence of such, so called, precursor events
(e.g., \citealt{Arnett+Meakin2011_TurbulentCells_precursor}; \citealt{Shiode+Quataert2014_WaveDrivenMassLoss}; \citealt{Fuller+2018_precursors}).
However, as discussed earlier, this scenario still requires an explanation for the lack of intermediate width emission
lines in the early spectrum of the event.

\subsection{Rates}

Estimating the rate of such events is difficult due to two reasons.
First, one has to define the threshold in term of luminosity and time scale we are interested in, which, in turn, requires
excellent observations of a large number of transients.
Second, one needs to know the discovery and followup efficiency of the survey for such events.
In the meanwhile, we can get an order-of-magnitude estimate for the rate by assuming that ZTF has found one object
in 3 years of observations, and that it covers about 1/10 of the sky at any given moment with high-enough cadence
to detect such events, and that it has a 20\% spectroscopic efficiency for faint transients.
Furthermore, assuming that AT\,2018lqh was found at a redshift that contains half the available volume for discovery, we estimate an order-of-magnitude rate of $\sim 10^{-7}$\,Mpc$^{-3}$\,yr$^{-1}$.
This rate is about two orders of magnitude lower than the supernovae rate, and roughly of the same order of magnitude
as the NS-NS merger rate (\citealt{Abbott+2017_GW170817_LIGO}).

\subsection{Discriminating between the models with future observations}

In the future, with additional observations, it will be possible to discriminate between the interaction-followed-by-cooling
scenario
and the radioactive-heating scenario.
Specifically, if the ejecta is powered by radioactivity, we would expect that after the ejecta
becomes optically thin, we will start to see broad features that correspond to the high velocities of the ejecta.
In contrast, such broad features are less likely in the interaction case.
Furthermore, such a scenario can be tested using early UV observations
(e.g., ULTRASAT; \citealt{Sagiv+2014_ULTRASAT}), which will better constrain the temperature and extinction.
In addition, for supernovae powered by CSM interaction,
we expect a relation between
the shock velocity, peak bolometric luminosity and
rise time (e.g., \citealt{Ofek+2014_IIn_rise_peak}).
Finally, such events are expected to be detected in X-ray observations (e.g.,
\citealt{Waxman+2007_GRB060218_ArelativisticShocjBreakout};
\citealt{Soderberg+2008_SN2008D_Xray};
\citealt{Svirski+2012_OpticalToXrays_WindBreakoutLightCurves}; \citealt{Svirski+2014_SpectrumLightCurves_WindBreakoutSN}; \citealt{Ofek+2013_IIn_X}; \citealt{Ofek+2014_NuSTAR}),
and radio (e.g., \citealt{Chevalier+Fransson1994_Emission_SN_CSM_Interaction}; \citealt{Chandra+2015_SN2010jl_Xrays_Radio}; \citealt{Ho+2019_AT2018cow_Radio}).

Finding such short-duration transients and obtaining followup observations is challenging.
One future survey that is designed for the detection of short-duration transients is the Large Array Survey Telescope (LAST; Ofek \& Ben-Ami 2020). LAST will use a large fraction of its observing time to conduct a fast cadence survey of $\gtorder2000$\,deg$^{2}$, eight times per night, with a limiting magnitude of 21.

\acknowledgments

E.O.O. is grateful for the support of
grants from the 
Willner Family Leadership Institute,
André Deloro Institute,
Paul and Tina Gardner,
Israel Science Foundation,
Minerva,
BSF, BSF-transformative,
Weizmann-UK,
and the I-CORE program of the Planning
and Budgeting Committee of the Israel Science Foundation (ISF).
AGY's research is supported by the EU via ERC grant No. 725161, the ISF GW Excellence Center, an IMOS space
infrastructure grant and BSF/Transformative, Minerva and GIF
grants, as well as The Benoziyo Endowment Fund for the
Advancement of Science, the Deloro Institute for Advanced
Research in Space and Optics, The Kimmel Center for
planetary science, The Veronika A. Rabl Physics Discretionary
Fund, Paul and Tina Gardner, Yeda-Sela and the WIS-CIT
joint research grant; AGY is the recipient of the Helen and
Martin Kimmel Award for Innovative Investigation.
Based on observations obtained with the  48-inch Samuel Oschin Telescope and  60-inch telescope at the Palomar Observatory as part of the Zwicky
Transient Facility project. ZTF is supported by the National Science Foundation under Grant No. AST-1440341 and a collaboration that includes Caltech, IPAC, Weizmann Institute of Science,  Oskar Klein Center at Stockholm University, University of Maryland, University of Washington, Deutsches Elektronen-Synchrotron and Humboldt University, Los Alamos National Laboratories, TANGO Consortium of Taiwan, University of Wisconsin at
Milwaukee, and Lawrence Berkeley National Laboratories. Operations are conducted by COO, IPAC, and UW.

\bibliography{papers.bib}
\bibliographystyle{aasjournal}

\end{document}